\documentclass[a4paper,fleqn]{cas-dc}

\usepackage{color}
\usepackage{cite}
\usepackage{multirow}
\usepackage{multicol}
\usepackage{hyperref}
\usepackage{array}
\usepackage{float}

\newlength{\bibsep}
\def\tsc#1{\csdef{#1}{\textsc{\lowercase{#1}}\xspace}}
\tsc{WGM}
\tsc{QE}
\tsc{EP}
\tsc{PMS}
\tsc{BEC}
\tsc{DE}

\begin{document}

\let\WriteBookmarks\relax
\def\floatpagepagefraction{1}
\def\textpagefraction{.001}

\shorttitle{
Etching-to-deposition transition in SiO$_2$/Si$_3$N$_4$ using CH$_x$F$_y$ ion-based plasma etching: An atomistic study with neural network potentials
}

\shortauthors{Hyungmin An et~al.}

\title [mode=title]{
Etching-to-deposition transition in SiO$_2$/Si$_3$N$_4$ using CH$_x$F$_y$ ion-based plasma etching: An atomistic study with neural network potentials
}                      



%
\author[1]{Hyungmin An}[
orcid=0000-0003-2529-4247
]
\fnmark[1]
\credit{Conceptualization, Investigation, Data curation, Validation, Formal analysis, Methodology, Software, Visualization, Writing – original draft}

\author[1]{Sangmin Oh}[
orcid=0000-0002-4229-3302
]
\fnmark[1]
\credit{Conceptualization, Investigation, Data curation, Validation, Formal analysis, Methodology, Software, Visualization, Writing – original draft}


\author[2]{Dongheon Lee}[
orcid=0000-0001-8774-2821
]
\credit{Conceptualization, Writing – review \& editing}
\author[2]{Jae-hyeon Ko}
\credit{Conceptualization, Writing – review \& editing}
\author[2]{Dongyean Oh}[
orcid=0000-0002-1262-1838
]
\credit{Conceptualization, Writing – review \& editing}

\author[1]{Changho Hong}[
orcid=0000-0001-8163-5443
]
\cormark[1]
\ead{mk01071@snu.ac.kr}
\credit{Conceptualization, Investigation, Methodology, Software, Writing – original draft, Supervision}

\author[1,3,4]{Seungwu Han}[
orcid=0000-0003-3958-0922
]
\cormark[1]
\ead{hansw@snu.ac.kr}
\credit{Conceptualization, Writing – review \& editing, Supervision, Project administration, Resources}

\affiliation[1]{
    organization={Department of Materials Science and Engineering, Seoul National University},
    postcode={08826},
    city={Seoul},
    country={Korea}}

\affiliation[2]{
    addressline={SK hynix Inc.}, 
    postcode={17336},
    city={Icheon-si, Gyeonggi-do},
    country={Korea}}

\affiliation[3]{
    organization={AI center},
    addressline={Korea Institute for Advanced Study}, 
    postcode={02455},
    city={Seoul},
    country={Korea}}

\affiliation[4]{
    organization={Research Institue of Advanced Materials, Seoul National University},
    postcode={08826},
    city={Seoul},
    country={Korea}}

\fntext[fn1]{H.A. and S.O. contributed equally.}
\cortext[cor1]{Corresponding author}

\begin{abstract}
Plasma etching, a critical process in semiconductor fabrication, utilizes hydrofluorocarbons both as etchants and as precursors for carbon film formation, where precise control over film growth is essential for achieving high SiO$_2$/Si$_3$N$_4$ selectivity and enabling atomic layer etching. 
In this work, we develop neural network potentials (NNPs) to gain atomistic insights into the surface evolution of SiO$_2$ and Si$_3$N$_4$ under hydrofluorocarbon ion bombardment. 
To efficiently sample diverse local configurations without exhaustive enumeration of ion-substrate combinations, we propose a vapor-to-surface sampling approach using high-temperature, low-density molecular dynamics simulations, supplemented with baseline reference structures. 
The NNPs, refined through iterative training, yield etching characteristics in MD simulations that show good agreement with experimental results.
Further analysis reveals distinct mechanisms of carbon layer formation in SiO$_2$ and Si$_3$N$_4$, driven by the higher volatility of carbon–oxygen byproducts in SiO$_2$ and the suppressed formation of volatile carbon–nitrogen species in Si$_3$N$_4$. 
This computational framework enables quantitative predictions of atomistic surface modifications under plasma exposure and provides a foundation for integration with multiscale process modeling, offering insights into semiconductor fabrication processes.
\end{abstract}

\begin{highlights}
\item Plasma etching of SiO$_2$ and Si$_3$N$_4$ is simulated using NNPs with various HFC ions.
\item In SiO$_2$, etching-to-deposition transition occurs via oxygen-deficient mixed layer formation.
\item In Si$_3$N$_4$, N is sputtered and carbon film forms under all conditions due to excess carbon.
\end{highlights}

\begin{keywords}
plasma etching \sep 
hydrofluorocarbon \sep 
density functional theory \sep 
neural network potential \sep 
silicon oxide \sep 
silicon nitride
\end{keywords}

\maketitle

\section{Introduction}

With the increasing global demand for computing resources, enhancing the integration density of semiconductor devices has become a critical challenge.
Plasma-assisted etching is essential for overcoming this challenge, supporting the fabrication of high-aspect-ratio (HAR) structures \cite{cao2023future}.
In plasma-assisted etching, high-energy ions accelerated by an electric field bombard the surface of the target material, transferring the lithographic pattern to the underlying layers of the integrated circuit.
This approach allows for directional anisotropic etching and material selective etching, making it particularly suitable for HAR semiconductor structures.
Among plasma-assisted etching methods, reactive ion etching (RIE) is widely utilized due to its ability to achieve high etch yields through the synergistic effects of chemical reactions and physical sputtering \cite{coburn1979ion,coburn1979plasma}. 
Additionally, atomic layer etching (ALE) is gaining attention for its precise etch depth control and high material selectivity \cite{kanarik2015overview,kanarik2018atomic,oehrlein2015atomic}. 

A central challenge in reactive ion etching (RIE) is achieving both high etch rates and material selectivity, particularly between widely used dielectric materials such as SiO$_2$ and Si$_3$N$_4$ \cite{hayashi1996characterization,kastenmeier1996chemical,schaepkens1999study,iijima1997highly,ito2002subsurface,standaert2004role,lee2010ultrahigh,kurihara2011measurements,miyawaki2012highly,miyake2014characterization,li2016fluorocarbon,lin2018achieving,cha2019low,kim2020plasma}.  
Hydrofluorocarbon (HFC) plasmas are widely used in this context, as they not only allow for flexible control over etch selectivity through chemical composition adjustments, but also facilitate the formation of carbon passivation layers that promote anisotropic etching profiles and suppress profile deformation issues such as bowing \cite{oehrlein2024future}.
A representative example is the selective etching of SiO$_2$ over Si$_3$N$_4$ using HFC gases \cite{schaepkens1999study}.  
At the etch front, SiO$_2$ readily reacts with fluorine-containing species to produce volatile byproducts such as SiF$_x$ and CO$_x$, resulting in efficient material removal. In contrast, Si$_3$N$_4$ shows lower reactivity toward volatile species.
Consequently, carbon species from the plasma are more likely to accumulate on the Si$_3$N$_4$ surface, forming a thicker carbon film that suppresses the etch rate.  
Such material-dependent carbon layer formation is a key factor governing etch selectivity in fluorocarbon-based plasma etching.  
Therefore, understanding how carbon films develop at the etch front as a function of substrate and plasma conditions is essential for optimizing RIE performance.

Accordingly, extensive experimental studies on SiO$_2$ and Si$_3$N$_4$ systems have investigated how feed gases influence the properties of carbon films and how these characteristics correlate with etching performance. For instance, the addition of H$_2$ into fluorocarbon plasmas scavenges fluorine by forming HF, increasing the C/F ratio of the carbon film and consequently reducing the etch rate \cite{fukasawa1994high,marra1997effect}. 
Similar outcomes have also been reported with CH$_4$ addition \cite{lee2010ultrahigh}. 
Furthermore, investigations employing various C$_x$F$_y$ gases have consistently demonstrated that conditions producing thinner carbon films typically yield lower C/F ratios and higher etch rates \cite{lee2010ultrahigh, li2002fluorocarbon, kim2020plasma, schaepkens1999study}.
Collectively, these findings support the conclusion that thicker carbon films with lower fluorine contents are correlated with reduced etch rates.

The relevance of carbon film formation has become particularly evident in atomic layer etching (ALE), a method gaining increasing attention in recent years. 
In ALE, the surface is modified firstly by depositing a controlled, nanometer-scale layer using fluorocarbon, followed by bombardment with energetic ions generated from Ar or O$_2$ plasma, which allows for highly controlled etching at an ultrathin scale \cite{li2016fluorocarbon,tsutsumi2017atomic,gasvoda2017surface,lin2018achieving,gasvoda2019surface,kim2020plasma}. 
In the ALE process, longer fluorocarbon exposure leads to greater surface modification, yet the substrate removal per cycle becomes self-limiting as etch time and film thickness increase \cite{tsutsumi2017atomic}.
To optimize process parameters such as HFC gas composition, the time of each step and bias potentials, precise control on the surface modification and carbon layer formation is necessary.
Nevertheless, due to the inherently complex nature of plasma environments, where ions and radicals coexist and act synergistically \cite{graves2024science}, isolating the specific role of ion bombardment in carbon film formation and etching dynamics remains challenging.

Ion beam experiments offer a controlled environment to isolate the effects of ion species, energy, and dose by irradiating surfaces with ions in the absence of radicals \cite{karahashi2014ion}. 
For SiO$_2$ and Si$_3$N$_4$ systems, previous studies have examined changes in surface composition and film thickness with ion dose \cite{ishikawa2003transitional,toyoda2004beam,yanai2005mass,ito2011hydrogen,bello1994importance}; variations in etch yield with ion energy \cite{toyoda2004beam,ito2011hydrogen,shibano1993etching,yamaguchi2000etching}; differences in yield across ion species \cite{karahashi2004etching,yanai2005mass,ito2011hydrogen,shibano1993etching}; and the impact of hydrogen on yield and surface chemistry \cite{ito2011hydrogen}. 
These studies have also reported carbon film formation under specific conditions: some observed deposition below a certain energy threshold \cite{shibano1993etching,yamaguchi2000etching}; others documented dose-dependent increases in film thickness and attributed the transition from initial etching to carbon deposition to fluorine assimilation mechanisms \cite{ishikawa2003transitional}; still others suggested that modified layer formation and associated changes in surface composition play a role in initiating film growth \cite{yanai2005mass,ito2011hydrogen}.
While these works offer plausible explanations for the onset of carbon film formation and the preceding etching behavior, they fall short of providing atomistic-level validation of the underlying mechanisms.
Understanding this dynamic behavior is particularly critical in ALE, where precise control over carbon dose and exposure time is essential to prevent excessive film accumulation that may cause etch stop or degrade etch selectivity \cite{hirata2020mechanism}.

Motivated by these atomic-scale considerations, molecular dynamics (MD) simulations have been widely employed to study carbon film formation during plasma etching of SiO$_2$ and Si$_3$N$_4$. 
Prior works have investigated various aspects of ion–surface interactions, such as film composition, etch yield, surface modification, and byproduct generation, using different ion species, energies, and substrate materials \cite{tanaka2000new,smirnov2005molecular,rauf2007molecular,miyake2014characterization,ito2014tight,cagomoc2022molecular,cagomoc2023molecular}. 
Although these studies have revealed general trends, including the effects of ion incidence angle and dose on surface composition and depth profiles, they predominantly rely on classical interatomic potentials, which require labor-intensive parameter fitting specific to each element or bonding environment \cite{muser2023interatomic,harrison2018review,martinez2013fitting}. 
This hinders scalability, as extending the model to new elements or chemistries often demands additional expertise and substantial refitting effort.
Furthermore, these classical models often yield byproduct distributions that deviate from experimental observations \cite{toyoda2004beam,cagomoc2022molecular}. 
Some studies have included hydrogen-containing ions \cite{miyake2014characterization}, but investigations of hydrogen effects remain scarce and have been limited to only a few specific systems.
On the other hand, first-principles methods such as density functional theory (DFT) can model interatomic interactions without empirical fitting and offer greater transferability, but their high computational cost restricts system size to a few hundred atoms and time scale to a few ps. 
These limitations highlight the need for an interatomic potential that combines DFT-level accuracy with the scalability of classical MD—enabling accurate simulations of dynamic film growth, byproduct formation, and material mixing in plasma etching environments.

In this respect, neural network potentials (NNPs)  present a promising alternative, offering high accuracy and computational efficiency suitable for large-scale simulations \cite{friederich2021machine,mortazavi2023atomistic,mortazavi2025recent,fedik2022extending,wan2024construction,wang2024machine}.
Nevertheless, generating comprehensive training datasets that accurately capture diverse atomic environments remains a major bottleneck in applying these models effectively \cite{10.1080/27660400.2023.2269948}.  
This issue is particularly severe in etching applications, where the local chemical and structural environments evolve rapidly and span a wide configurational space. 
Some of the authors previously developed a neural network potential (NNP) for Si$_3$N$_4$/HF etching by integrating multiple training set types: (1) baseline structures such as bulk and slab configurations, (2) reaction-specific datasets designed to capture key bond-breaking events relevant to gas–surface interactions, (3) general-purpose configurations obtained via high-temperature molecular dynamics to sample low-coordinated and disordered states, and (4) iterative refinement cycles to improve accuracy \cite{hong2024atomistic}. 
In that study, the final NNP was rigorously validated against DFT calculations and it was demonstrated that the bottom-up mesoscale model informed by microscopic simulations reasonably predicts surface composition evolution.
Yet, extending this framework to more complex systems presents fundamental challenges. 
In particular, when carbon-containing ions are introduced—as is common in plasma etching—surface dynamics are no longer limited to etching. 
Instead, carbon film formation and the emergence of interfacial mixed layers must also be captured \cite{gasvoda2020etch}. 
These phenomena involve a wider variety of atomic environments and surface states, greatly expanding the space of possible bond-breaking and bond-forming reactions. 
As a result, the required training set becomes significantly more complex and difficult to design manually, increasing the demand for both computational resources and domain-specific expertise.

In this study, we propose a more scalable and generalizable approach by avoiding the need for reaction-specific datasets. 
Specifically, we construct a training dataset that systematically samples a broad range of compositions and atomic environments including substrate regions, mixed layers, and carbon films, using both bulk-density and low-density simulations coined as vapor-to-surface (VtS) sampling.
This density modulation allows for enhanced sampling of low-coordinated configurations, enabling effective coverage of diverse surface chemistries without prior assumptions. 
Our method thus aims to replace expert-driven dataset construction with an automated, data-rich sampling strategy better suited for chemically and structurally complex systems.
We validate the resulting NNPs against experimental data by comparing etch yields, film thicknesses, and surface composition profiles, all of which show close agreement.
We then apply the trained NNPs to analyze carbon film formation across different ion species and energy conditions.
In the SiO$_2$ system, we find that an initial excess-carbon layer creates an oxygen-deficient surface, which subsequently transforms into a fluorocarbon film through Si etching.
For the Si$_3$N$_4$ system, fluorocarbon film formation is observed across all ion–energy combinations, accompanied by preferential sputtering of nitrogen.
We further compare how the number of F and H atoms in the incident ion, as well as the ion energy, influence the formation and characteristics of the mixed layer.

\begin{figure*}[htbp]
\centering
\includegraphics[width=1.0\linewidth]{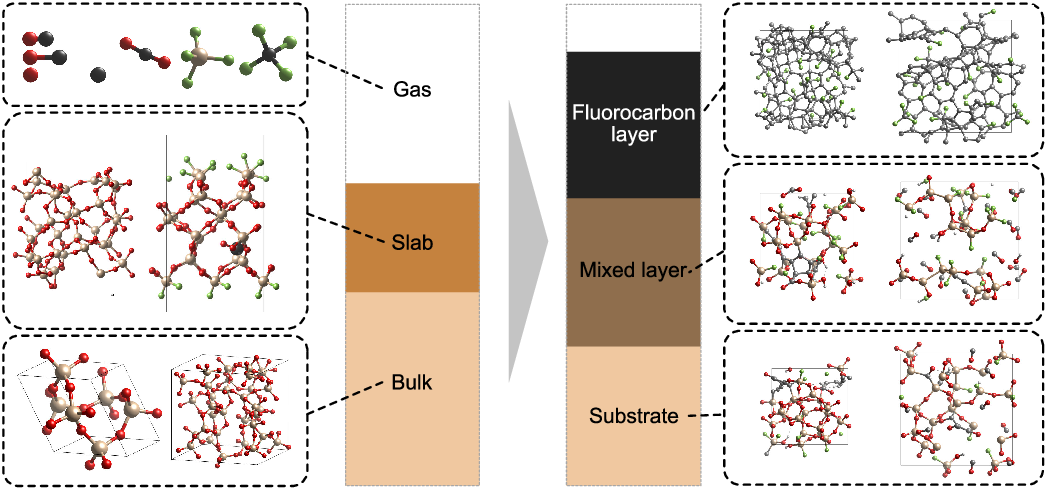}
\caption{Schematic overview of the training set used to develop the primitive NNP in this study.
(Left) The dataset includes bulk and slab structures that represent the substrate before etching, as well as various gaseous species that describe etching ions and byproducts.
(Right) It also contains a wide range of structures representing configurations during etching, including various compositions and both high- and low-density structures corresponding to the mixed layer and the fluorocarbon layer.} \label{fig:2_1_1_trainingset_overview}
\end{figure*}

\section{Computational methods}
\subsection{Training protocol}
Our training procedure follows the same overall structure as our previous work \cite{hong2024atomistic}: we first construct a primitive NNP using baseline structures and general-purpose datasets, and then refine it through iterative learning based on etching MD simulations in small cells. 
The key difference in this study is that we omit reaction-specific datasets entirely and instead enrich the general-purpose dataset by expanding the composition and density diversity of the general-purpose dataset using vapor-to-surface (VtS) sampling. 
This strategy allows us to sample a broader range of low-coordination and surface-like configurations without manually designing reaction pathways.
The following sections are organized into two parts. 
First, we describe the construction of the primitive NNP, including the baseline structures and the general-purpose dataset. 
Second, we outline the iterative learning procedure used to improve accuracy in relevant regions of the configuration space.
We develop separate NNPs for the SiO$_2$ and Si$_3$N$_4$ systems, each trained to simulate interactions with CH$_x$F$_y$ ions. 

\subsubsection{Primitive NNP}

\begin{table}[width=.9\linewidth,h!]
\centering
\caption{Summary of the training subsets used to construct the primitive NNP. Each entry lists a distinct subset, its representative chemical composition, and the number of atomic configurations included. Subsets not involving oxygen or nitrogen are shared between the SiO$_2$ and Si$_3$N$_4$ systems.}
\begin{tabular}{ c c r r }
 \hline
 & Description & \multicolumn{2}{c}{Training points} \\ 
 Type & (Si:O/N:C:H:F) & SiO$_2$ & Si$_3$N$_4$ \\ 
 \hline
 bulk & distorted crystalline     &     4 656 &   4 228 \\
      & crystalline MD            &    18 000 &  56 000 \\
      & amorphous                 &   122 364 & 106 112 \\
 \hline
 slab & crystalline               &    32 800 &  21 600 \\
      & amorphous                 &   111 375 & 126 000 \\
      & interstitial defect       &    83 000 &  54 500 \\
      & vacancy defect            &    64 800 &  42 693 \\
      & gas channeling            &    10 080 &  37 080 \\
 \hline
 gas  & dimers                    &     1 642 &   1 410 \\
      & polyatomic gases          &     6 200 &   2 556 \\
 \hline
substrate            & 3:6:3:1:3            & 74 496     & .          \\
dominant             & 3:3:3:1:3            & .          & 60 684     \\
                     & 3:6:2:1:1            & 70 616     & .          \\
                     & 3:4:2:1:1            & .          & 59 906     \\
                     & 6:12:1:1:1           & 81 480     & .          \\
                     & 4:10:1:1:1           & .          & 66 130     \\
                     & 3:3:1:1:1            & 76 725     & .          \\
                     & 6:8:1:1:1            & .          & 55 505     \\
\hline
mixed-layer          & 1:2:8:1:6            & 83 808     & .          \\
                     & 1:1:8:1:6            & .          & 51 765     \\
                     & 1:1:1:12:1           & 74 496     & .          \\
                     & 1:1:1:13:1           & .          & 66 130     \\
                     & 1:8:1:1:1            & 74 208     & .          \\
                     & 1:1:0:0:3            & 77 700     & .          \\
                     & 7:1:7:1:1            & .          & 66 130     \\
\hline
carbon-film          & 0:0:1:0:0.2          & 47 232     & 93 888     \\
(bulk)               & 0:0:1:0:0.4          & 39 168     & 77 760     \\
                     & 0:0:1:0.05:1         & 64 224     & 127 872    \\
                     & 0:0:1:0.15:1         & 64 224     & 127 872    \\
                     & 0:0:1:0.1:0.1        & 64 224     & 127 872    \\
                     & 0:0:1:0.1:0.4        & 64 224     & 127 872    \\
                     & 0:0:1:0.1:0.6        & 64 224     & 127 872    \\
                     & 0:0:1:0.1:0.8        & 64 224     & 127 872    \\
\hline
carbon-film          & 0:0:1:0:0.2          & \multicolumn{2}{c}{28 800} \\
(slab)               & 0:0:1:0:0.4          & \multicolumn{2}{c}{28 800} \\
                     & 0:0:1:0.05:1         & \multicolumn{2}{c}{27 360} \\
                     & 0:0:1:0.15:1         & \multicolumn{2}{c}{27 360} \\
                     & 0:0:1:0.1:0.1        & \multicolumn{2}{c}{28 800} \\
                     & 0:0:1:0.1:0.4        & \multicolumn{2}{c}{28 800} \\
                     & 0:0:1:0.1:0.6        & \multicolumn{2}{c}{27 360} \\
                     & 0:0:1:0.1:0.8        & \multicolumn{2}{c}{27 360} \\
\hline
\end{tabular}
\label{table:training_set}
\end{table}

\autoref{fig:2_1_1_trainingset_overview} presents a schematic illustration of the training set for constructing the primitive NNP, while \autoref{table:training_set} summarizes the types of training subsets and the number of configurations used.
The primitive NNP is built using baseline structures and general-purpose datasets. 
The baseline structures include three types: (1) bulk, (2) slab, and (3) gas-phase configurations. 
Datasets that do not contain oxygen or nitrogen are shared across the SiO$_2$ and Si$_3$N$_4$ systems.

The bulk structures used in this study can be categorized into three types: (1) distorted crystalline structures, (2) crystalline MD structures, and (3) amorphous structures.
Distorted crystalline structures are generated by applying normal and shear strains for structure-optimized crystalline unit cells, from -0.1 to +0.1 using the \texttt{Pymatgen} package \cite{ong2013python}. 
For the SiO$_2$ system, four crystalline SiO$_2$ phases ($\alpha$-quartz, $\beta$-quartz, $\alpha$-tridymite, $\beta$-cristobalite) unit cells are used. 
For the Si$_3$N$_4$ system, two crystalline Si$_3$N$_4$ phases ($\alpha$, $\beta$) unit cells are used.
Crystalline MD structures are generated by NVT-MD simulations (600 K, 10 ps) using the $2\times2\times2$ supercells of crystalline unit cells. 
For the SiO$_2$ and Si$_3$N$_4$ systems, two crystalline phases are used for each: $\alpha$-quartz and $\beta$-quartz for SiO$_2$, and $\alpha$-phase and $\beta$-phase for Si$_3$N$_4$.
Amorphous structures are initially generated by randomly spraying atoms within a cell with the fixed volume, followed by a NVT-MD (5000 K, 2 ps). 
This is followed by a sequential melt–quench–anneal process: melting (4000 K, 20 ps), quenching (from 4000 K to 300 K at a rate of –100 K/ps), and annealing (500 K, 15 ps). 
For the SiO$_2$ system, two different target densities are considered for 99-atom cell: 2.34 g/cm$^3$ and 2.2 g/cm$^3$. 
For the Si$_3$N$_4$ system, target density of 3.1 g/cm$^3$ is used for 112-atom cell; also, the melting temperature of 3000 K is used.

The slab structures used in this study consist of five categories: (1) crystalline structures, (2) amorphous structures, (3) interstitial defect structures, (4) vacancy defect structures, and (5) structures with gas channeling.
Crystalline structures are constructed by adding vacuum along the [001] direction to the $2\times2\times2$ supercell of bulk crystalline unit cells. 
Subsequently, NVT-MD (1 ps) simulations are conducted for two different temperatures of 300 K and 2000 K each. 
For the SiO$_2$ system, the $\alpha$-quartz phase is used with two terminations: Si-terminated surfaces passivated with fluorine, and O-terminated surfaces passivated with hydrogen. 
For the Si$_3$N$_4$ system, the $\alpha$-phase slab is used, with nitrogen atoms passivated with hydrogen and silicon atoms passivated with fluorine.
Amorphous slab structures are prepared by adding vacuum to the previously generated amorphous bulk structures. 
The system then undergoes sequential NVT-MD simulations comprising annealing (500 K, 10 ps), melting (from 300 K to 3000 K, 10 ps), and maintaining liquid state (3000 K, 20 ps).
Interstitial defect structures are generated by introducing a single interstitial atom (Si, O/N, C, H, or F) at the center of the slab region. 
These systems then undergo NVT-MD (2000 K, 8 ps).
Vacancy defect structures are constructed by introducing atomic vacancies into crystalline slab structures, followed by NVT-MD simulations (2000 K, 4 ps).
For the SiO$_2$ system, two distinct monovacancy sites are considered for each termination, with separate structures generated for Si and O vacancies at each site.  
For the Si$_3$N$_4$ system, Si and N monovacancies are similarly introduced at multiple sites.
Structures with gas channeling follow the approach used in our previous study \cite{hong2024atomistic}, where NVE-MD simulations are performed to model the movement of HF molecules through channel-like paths within crystalline slabs along the (0001) direction. 
For both the SiO$_2$ and Si$_3$N$_4$ systems, HF molecules are initially placed at the center of the hole in the slab with a kinetic energy of 10 eV, and their trajectories are simulated under identical conditions for both terminations.

The gas-phase structures used in this study consist of two categories: (1) dimer structures and (2) polyatomic gas structures.
The dimer structures are constructed from all possible unique pairs among the five constituent elements in each system—Si, C, H, F, and either O or N—resulting in 15 distinct combinations per system. 
For each dimer, the interatomic distance is varied from –10\% to +30\% relative to the equilibrium bond length, in increments of 1\%p.
The polyatomic gas structures are built by placing two to four surrounding atoms—selected from H, F, O, and N, i.e., excluding Si and C—around a central Si or C atom. 
All possible element combinations are enumerated within this constraint. 
The atomic geometries are chosen based on the number of surrounding atoms—linear, triangular, or tetrahedral. 
After structure optimization, unstable molecules (i.e., those with broken bonds) are discarded. 
The remaining molecules undergo NVT-MD (1000 K, 2 ps).

The general-purpose training sets are constructed to cover diverse atomic configurations relevant to plasma etching of SiO$_2$/Si$_3$N$_4$ using hydrofluorocarbons. 
This diversity is categorized based on composition into three representative regimes commonly observed during the etching process \cite{karahashi2004etching,ito2011hydrogen}: (1) the substrate-dominant regime, where Si and O/N atoms are prevalent; (2) the mixed-layer regime, where Si, O/N and C, H, F atoms coexist in comparable amounts; and (3) the fluorocarbon film regime, where C, H, F atoms dominate.

To construct training data for the substrate-dominant and mixed-layer regimes, we generate bulk structures specific to SiO$_2$ and Si$_3$N$_4$ using melt–quench–anneal simulations across a range of selected atomic ratios. The simulation protocol is identical to that described previously for baseline bulk structures. 
The target density for each composition is determined by linear interpolation between the crystalline density of the substrate and the density of Teflon (2.2 g/cm$^3$) \cite{schaepkens1998selective}, according to the relative fraction of substrate and fluorocarbon components. Specifically, we use 2.6 g/cm$^3$ for SiO$_2$ and 3.17 g/cm$^3$ for Si$_3$N$_4$ as the reference crystalline densities.

For the fluorocarbon-rich regime, we construct a bulk dataset composed of eight different compositions.
Here, we use a fixed density of 3.2 g/cm$^3$ and employ a slightly different melt–quench–anneal scheme, following the previous DFT-MD approach to make amorphous carbon structures \cite{park2020influence}: melting (5000 K, 10 ps), quenching (from 5000 K to 0 K at a rate of –278 K/ps), and annealing (500 K, 15 ps).
In addition to the bulk configurations, we include slab structures, which are generated by adding a vacuum region above the bulk amorphous structures. 
These slab structures then undergo sequential NVT-MD simulations: anneal (500 K, 10 ps) and melt (4500 K, 20 ps).

To further enhance configurational diversity, particularly for undercoordinated and surface-like motifs, we incorporate a low-density variant—referred to as vapor-to-surface (VtS) sampling—across all three composition regimes. 
For the substrate-dominant and mixed-layer regimes, the VtS structures are generated by using half the density of the corresponding normal-density bulk structures. 
For the fluorocarbon film regime, the VtS structures are generated to sample environments containing a higher proportion of low-coordinated carbon atoms, at a fixed density of 2.3 g/cm$^3$ which is consistent with experimentally measured values for amorphous carbon \cite{iwaki2002estimation,lacerda2001hard}.

\begin{figure}[htbp]
\centering
\includegraphics[width=1.0\linewidth]{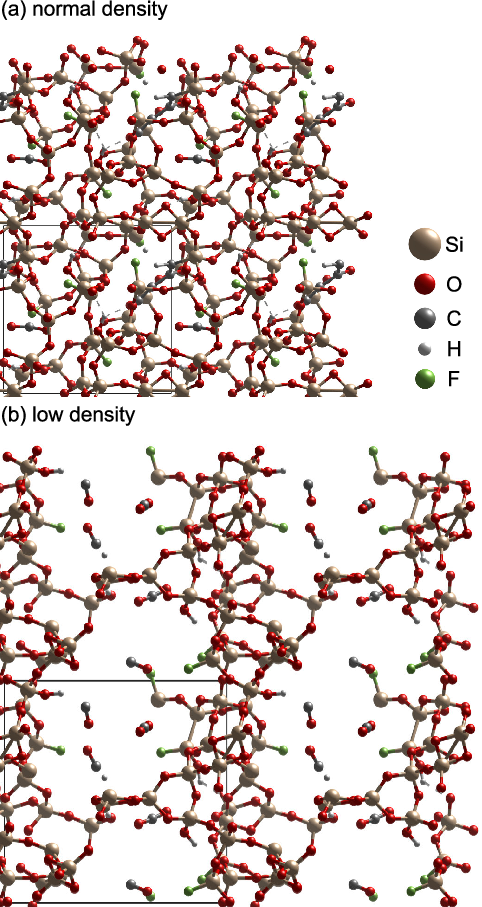}
\caption{Atomic configurations generated via the melt–quench–anneal process with the same composition (Si:O:C:H:F = 6:8:1:1:1) at two different densities: (a) normal density of 2.54 g cm$^{-3}$ and (b) low density of 1.27 g cm$^{-3}$. The solid black lines indicate the unit cell, and each configuration is shown with a replication factor of 2. Compared to the normal-density case, the low-density structure exhibits surface-like features surrounded by gaseous species.} \label{fig:2_1_1_vts_effect}
\end{figure}

These low-density simulations are effective in generating configurations rarely encountered under melt–quench–anneal conditions with bulk amorphous densities and share a conceptual similarity with nanoreactor-based approaches that explore alternative structural phases by modulating system density \cite{zhang2024exploring}. 
At reduced density, atoms either cluster to form surface-like structures or remain isolated as gas-phase species.
\autoref{fig:2_1_1_vts_effect} illustrates the structural differences between amorphous structures generated at normal and low densities.
Figure~S1 shows the distributions of fractional coordination numbers for each element in two datasets with identical compositions but different densities. 
For Si and C specifically, the distributions in the low-density set are clearly shifted toward lower coordination numbers, indicating the increased prevalence of undercoordinated environments.

\subsubsection{Iterative learning}
After training the primitive NNP, we apply an iterative learning strategy to further refine the model for etching MD simulations. 
First, amorphous slab structures of SiO$_2$ and Si$_3$N$_4$ are generated using NNP-based melt–quench-anneal simulations, followed by structural optimizations.  
To ensure that DFT single-point calculations remain computationally feasible, the initial slabs are constructed to contain approximately 100 atoms for both systems.
NNP-MD etching simulations are then performed by randomly placing and orienting ions above the slab surface. 
Each ion is injected at a fixed energy and a random incident angle. Although ions in plasma are charged, it is well established that charge neutralization occurs near solid surfaces prior to impact, due to processes such as Auger neutralization \cite{tully1977neutralization,pretzer1966ion}. As the neutralization assumption is common in similar computational studies \cite{ohta2001molecular}, we simulate neutral species with initial velocities corresponding to the specified kinetic energies.
The simulation consists of an initial 2 ps NVE-MD phase, followed by NVT-MD at 300 K, and continues until the slab temperature reaches 350 K. 
This process is repeated 50 times for each condition.
We consider two ion energies (10 eV and 30 eV) and four ion species (CF$^+$, CF$_3^+$, CH$_2$F$^+$, and CHF$_2^+$). 
For each ion impact, five snapshots are extracted at 0 ps (initial), 0.8 ps, 1.6 ps, 2.4 ps, and 4 ps, yielding a total of 2,000 configurations. 
DFT single-point calculations are performed for these snapshots, and the resulting data are added to the training set for model retraining.
This iterative learning cycle is repeated three times, resulting in the cumulative addition of 6,000 new configurations. 
All calculations using NNPs are performed with \texttt{LAMMPS} package \cite{thompson2022lammps}.

\subsubsection{Training settings}

The hybrid approach of combining NNP with short-range interatomic potential has been employed in several prior studies to improve accuracy and stability in highly repulsive regimes \cite{wood2019data,wang2022machine,byggmastar2019machine,fellman2025fast,liu2023large,thompson2015spectral,dubois2024atomistic,wang2019deep,sikorski2023machine}.
Our previous work also adopted this scheme for plasma–surface interactions \cite{hong2024atomistic}, and we briefly summarize the method here.
When calculating configurations involving atomic collisions at very short distances, extremely high energy and force values must be accurately captured.
However, such collision events can occur across a wide variety of local environments—including bulk phases, surfaces, and transient reaction intermediates—making it difficult to comprehensively sample all relevant configurations in the training set. 
We replace these high-energy interactions with the Ziegler-Biersack-Littmark (ZBL) potential \cite{ziegler1985stopping}, allowing for a hybrid approach that combines NNP with ZBL. 
The NNP is trained to learn only the DFT data with the ZBL potential subtracted, as \autoref{eqn:hybrid approach}. 
The ZBL cutoff distances we used are listed in Table~S1.

\begin{equation} \label{eqn:hybrid approach}
    E_{\text{NNP}}=E_{\text{DFT}}-E_{\text{ZBL}}
\end{equation}

We use \texttt{SIMPLE-NN} \cite{lee2019simple} package for training our NNP and running NNP-MD simulations. For training, we use 182 symmetry functions (32 radial, 150 angular) for Si, O, N, and C atoms, and 110 symmetry functions (20 radial, 90 angular) for H and F atoms. 
The cutoff radius for the radial symmetry functions is set to 4.5 Å when H or F atoms are involved, and 6.0 Å otherwise, while a uniform cutoff of 4.5 Å is used for all angular symmetry functions.
In preprocessing, each symmetry function undergoes min/max scaling and principal component analysis (PCA) whitening is subsequently applied. The training-validation split ratio is set to 9:1. The neural network architecture consists of two hidden layers with 60 nodes each, using the tanh activation function. 
Training is performed using the Adam optimizer, with an initial learning rate of 10$^{-4}$ for the first 200 epochs, which is later reduced to 10$^{-5}$ for the following 400 epochs. 
The loss function incorporates energy, force, and stress terms, with respective weighting coefficients of 1, 0.1, and 10$^{-6}$. 
After 600 epochs of training, the model with the lowest root-mean-squared-error (RMSE) is selected as the primitive NNP.

\subsection{DFT calculation settings}
We perform all DFT calculations using the Vienna \textit{Ab initio} Simulation Package (\texttt{VASP}) \cite{kresse1996efficiency}. For the exchange-correlation functional, we employ the Perdew-Burke-Ernzerhof (PBE) functional \cite{perdew1996generalized}. 
For the \textit{ab initio} MD simulations used in training, we employ soft pseudopotentials to reduce computational cost with spin-unpolarized calculations. The default cutoff energy defined in the soft pseudopotential for each element is used, and only the $\Gamma$-point is used for Brillouin zone sampling. For NVT-MD, temperature is controlled by the Nos\'{e}--Hoover thermostat. The timestep is set to 2 fs, and to ensure simulation stability at this timestep, the mass of hydrogen atoms is arbitrarily increased by a factor of three when present. After the initial sampling, single-point DFT calculations are conducted using the standard version of the pseudopotentials with spin polarized calculations. For these calculations, we apply a cutoff energy of 500 eV to ensure sufficient accuracy. We used k-point grid that ensures energy convergence of 10 meV/atom for bulk crystalline structures, otherwise $\Gamma$-point is used.

\begin{table}[width=.9\linewidth,h!]
\centering
\caption{List of molecular clusters removed after each ion incidence during etching MD simulations. These molecules are identified via graph-based clustering, and any cluster matching the composition listed here is removed. Entries labeled as “both” indicate that the molecule can appear in both SiO$_2$ and Si$_3$N$_4$ systems.}
\begin{tabular}{ m{5em}  m{3em}  m{12em} }
 \hline
 Type & System & Molecules \\
 \hline
 SiH$_x$F$_y$ & both & SiF$_4$, SiHF$_3$, SiH$_2$F$_2$, SiH$_3$F, SiH$_4$, SiF$_2$ \\
 \hline
 CH$_x$F$_y$  & both & CF$_4$, CHF$_3$, CH$_2$F$_2$, CH$_3$F, CH$_4$, CF$_2$ \\
 \hline
 \multirow{3}{4em}{Stable dimers} & both & H$_2$, F$_2$, HF \\
               & SiO$_2$ & O$_2$, CO \\
               & Si$_3$N$_4$ & N$_2$ \\
 \hline
 \multirow{2}{4em}{Byproducts}    & SiO$_2$ & H$_2$O, OF$_2$, OHF, CH$_2$O, CHFO, COF$_2$, CO$_2$ \\
               & Si$_3$N$_4$ & FCN, HCN, NH$_3$, NH$_2$F, NHF$_2$, NF$_3$ \\
 \hline
\end{tabular}
\label{table:molecule_to_erase}
\end{table}

\subsection{Etching simulation}
\subsubsection{MD settings}
After completing iterative learning, we perform large-scale etching MD simulations using the final NNP to compare with ion beam experiments. 
To minimize finite-size effects, amorphous slab structures of SiO$_2$ and Si$_3$N$_4$ are constructed with in-plane dimensions of 3 nm × 3 nm and a thickness of 6 nm. The simulation cell size along the z-axis is fixed at 20 nm to ensure sufficient vacuum spacing and the bottom
layer with a thickness of 6 Å is fixed.
To account for molecule–surface interactions, we include van der Waals (vdW) forces using the DFT-D2 formalism \cite{grimme2006semiempirical}, with a cutoff distance of 15 Å.
Each ion is injected at normal incidence with a fixed kinetic energy. 
The simulation protocol consists of a 2 ps NVE-MD phase followed by NVT-MD at 300 K. 
To enhance numerical stability during high-energy collisions, a variable timestep scheme \cite{nordlund1995molecular} is employed. 
The timestep is adaptively adjusted to satisfy three conditions: (1) the maximum atomic displacement per step does not exceed 0.1 Å, (2) the kinetic energy transferred to any atom is limited to 1 eV, and (3) the timestep does not exceed the initial value of 2 fs. 
During NVT-MD, a Langevin thermostat \cite{allen2017computer} is used to maintain the target temperature, and the simulation is terminated once the system stabilizes at 300 K.
We consider five ion species (CF$^+$, CF$_2^+$, CF$_3^+$, CH$_2$F$^+$, and CHF$_2^+$) and four incident energies (250, 500, 750, and 1000 eV). Additional energies (10, 25, 50, 75, 100, and 300 eV) are also used for selected species. 

\subsubsection{Byproducts and slab modification}
We assume that a sufficiently long time interval exists between ion impacts, allowing byproducts formed on the surface to fully evaporate. 
Such assumption has also been adopted in previous computational studies \cite{cagomoc2022molecular}.
Based on this assumption, after each ion impact, the atomic structure is analyzed to identify and remove reaction products. 
Bond networks are constructed by defining bonds as atom pairs with interatomic distances less than 1.3 times the equilibrium bond length of the corresponding dimer. 
Atoms connected via bonds are grouped into clusters, and clusters corresponding to predefined byproducts (\autoref{table:molecule_to_erase})—including experimentally known stable species \cite{mogab1978plasma,miyata1997absolute,toyoda2004beam}—are removed. 
Additionally, any cluster located more than 1 nm above the slab surface is eliminated, regardless of its composition or correspondence to predefined byproducts.

Following byproduct removal, the slab composition is evaluated to determine whether structural modification is necessary for every 10 ion incidences.
When deposition processes dominate, atoms are removed from the bottom of the slab to reduce computational cost without affecting surface dynamics. 
Conversely, when etching processes predominate, additional atoms from bulk amorphous structures are added to the slab bottom to prevent complete penetration of incoming ions.
To classify the system as being in a deposition regime, we calculate the combined fraction of carbon, hydrogen, and fluorine atoms located more than 6 Å above the slab bottom. 
If this fraction exceeds 70\%, the system is considered to be in a deposition regime, and atoms within 6 Å from the bottom are removed. 
If the fraction is below 70\%, we examine the lowest z-coordinate among all C, H, and F atoms. 
If any of these atoms are located within 12 Å of the slab bottom, the system is classified as undergoing etching, and 6 Å of material is appended to the bottom of the slab. 
These additional atoms are extracted from the original bulk amorphous structure to ensure compositional consistency.


\section{Results and discussion}
\subsection{Validation of NNP}
To investigate the mechanism of carbon film formation, it is essential that our simulations accurately reproduce the results of ion beam experiments. This requires verifying that the trained NNP can capture both steady-state and transient surface behaviors.
We first validate the NNP by comparing its predicted energies and forces with DFT results from the training set as well as reaction energies between snapshots from NNP MD trajectories, ensuring consistency with the DFT potential energy surface.
Next, we evaluate its ability to reproduce key experimental observables: etch yield serves as a steady-state property reflecting long-term material removal, while surface height and composition profiles capture non-steady-state responses. 

\subsubsection{Training results}
The final NNPs achieve energy and force mean-absolute-error (MAE) of 15.41 meV/atom and 0.285 eV/Å for the SiO$_2$ system, and 18.41 meV/atom and 0.358 eV/Å for the Si$_3$N$_4$ system, respectively. 
The parity plots are provided in Figure~S2 and training results for each subset of the training set are provided in Table~S2.
Table~S3 shows that a significant reduction in energy MAE is observed before and after iterative learning, which helps preventing energy shifts and maintaining small energy error for the subsequent MD simulations.
The energy differences for the entire trajectory in iterative learning are plotted in Figure~S3.

To assess whether the NNP accurately represents energy changes associated with bond breaking and formation, we compared reaction energies defined as the differences in total energy between snapshots obtained from MD simulations. Unlike absolute total energy errors, which can be masked by thermal vibrations, reaction energies explicitly measure the energy variations corresponding to chemically meaningful transformations.
Configurations are sampled using the same protocol as the iterative learning stage, where NNP-based etching MD simulations are performed on amorphous slabs with 20 ion impacts.
Three ion species (CF$^+$, CF$_3^+$, CH$_2$F$^+$) and two energies (20 and 50 eV) are used. 
From each trajectory, we extract snapshots at 10\% and 60\% of the total time sequence and optimize them using both DFT and NNP.
To focus on chemically distinct reactions, connectivity-based filtering was applied to remove duplicate structures, namely, configurations having identical bond connectivity despite differing atomic positions.
Figure~S4 shows strong agreement between DFT and NNP reaction energies, with R$^2$ values of 0.87 for the SiO$_2$ system and 0.80 for the Si$_3$N$_4$ system, demonstrating the reliability of the model across diverse reaction pathways and ion conditions.

\begin{figure*}[htbp]
\centering
\includegraphics[width=1.0\linewidth]{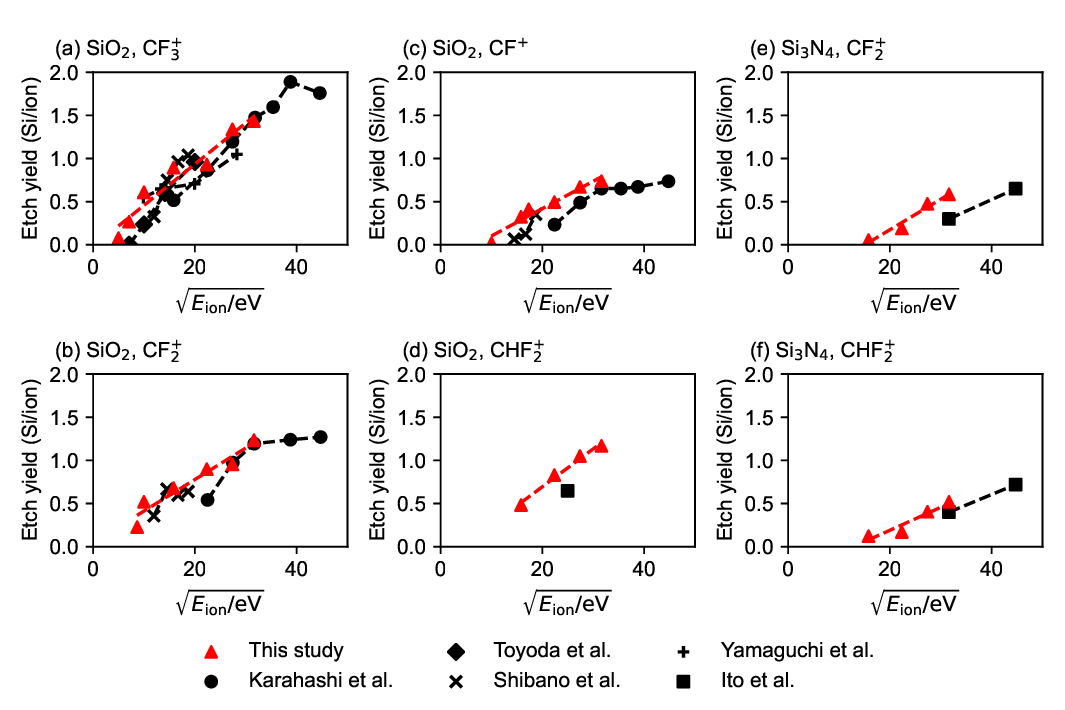}
\caption{Etch yield as a function of ion incidence energy for different combinations of substrate type and ion species. 
Red triangles represent simulation results from this work, while the other markers denote data from previous experimental ion beam studies \cite{karahashi2004etching,toyoda2004beam,shibano1993etching,yamaguchi2000etching,ito2011hydrogen}.
Dashed lines for the experimental data connect the reported points, whereas the dashed lines for this study indicate trend lines fitted to simulation points.} \label{fig:3_1_2_valid_etchyield}
\end{figure*}

\subsubsection{Etch yield}
To compare our etching MD results with experimental data, we compute the etch yield, which is defined as the number of Si atoms removed per ion incidence. 
In the previous experimental ion beam study, the etch yield did not depend on ion dose from 1.5 × 10$^{16}$ cm$^{-2}$ to 5 × 10$^{17}$ cm$^{-2}$ \cite{karahashi2004etching}. 
For direct comparison with ion beam experiments, we use an ion dose of 5 × 10$^{16}$ cm$^{-2}$, corresponding to 4,500 ion incidences on the 3 × 3 nm$^2$ lateral area of our simulation cell. 
At each point, the etch yield is averaged over the preceding 0.44 × 10$^{16}$ cm$^{-2}$ ion dose (corresponding to 400 ion incidences) to smooth out fluctuations. 
Figure~S5 shows that the etch yield converges within ion dose of 5 × 10$^{16}$ cm$^{-2}$ in most cases.

\autoref{fig:3_1_2_valid_etchyield} compares our etch yield values with experimental data \cite{toyoda2004beam,karahashi2004etching,ito2011hydrogen,shibano1993etching,yamaguchi2000etching}. 
Our results closely follow the universal behavior in which etch yield scales with the square root of ion energy \cite{steinbruchel1989universal}, which confirms the reliability of our NNP and MD protocol.
For the SiO$_2$ system, however, we observe a slight overestimation of etch yield at low energies within the given ion dose of 5 × 10$^{16}$ cm$^{-2}$, particularly for CF$^+$ ions below 500 eV.
As shown in Figure~S6, the etch yield for CF$^+$ at 300 eV gradually decreases and converges to zero at an extended dose of 1.5 × 10$^{17}$ cm$^{-2}$, indicating a delayed onset of carbon film deposition compared to experiments.
This behavior suggests that the transition from etching to deposition occurs more slowly in our simulations, and we further relate it with the removal height of the byproducts in the following section.

For the Si$_3$N$_4$ system, the etching yields from the simulations follow the experimental trends for CF$_2^+$ and CHF$_2^+$ ions. 
Although experimental data for CF$^+$ and CF$_3^+$ ions also exist for the Si$_3$N$_4$ system \cite{yanai2005mass}, we do not include them in the comparison due to a significant difference in SiN density (2.05 g/cm$^3$) in the experiment versus ours (3.1 g/cm$^3$).

\begin{figure*}[htbp]
\centering
\includegraphics[width=1.0\linewidth]{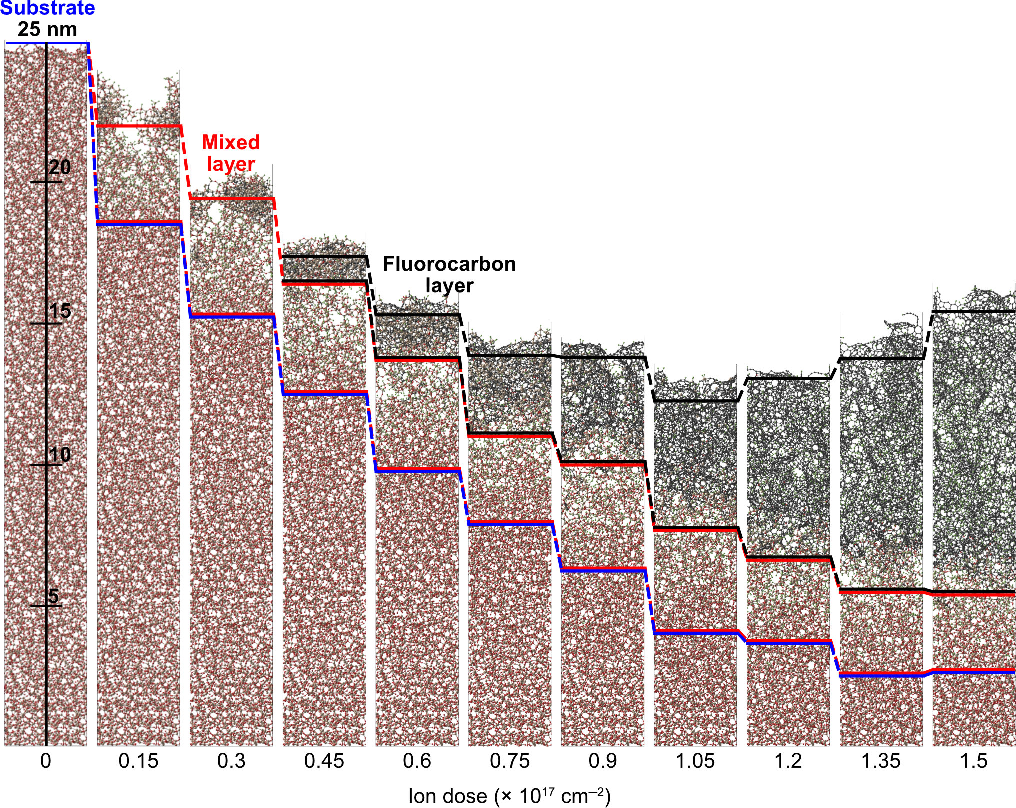}
\caption{Atomic configuration snapshots during etching MD with CF$^+$ ions at 300 eV on a SiO$_2$ substrate, shown as a function of ion dose. 
Each frame is generated by superimposing added and removed atomic structures relative to the initial configuration. 
The region below the blue line is classified as the substrate region, the region between the red lines as the mixed layer, and the region between the black lines as the fluorocarbon layer.
} \label{fig:3_1_3_valid_md_snapshot}
\end{figure*}

\begin{figure}[htbp]
\centering
\includegraphics[width=1.0\linewidth]{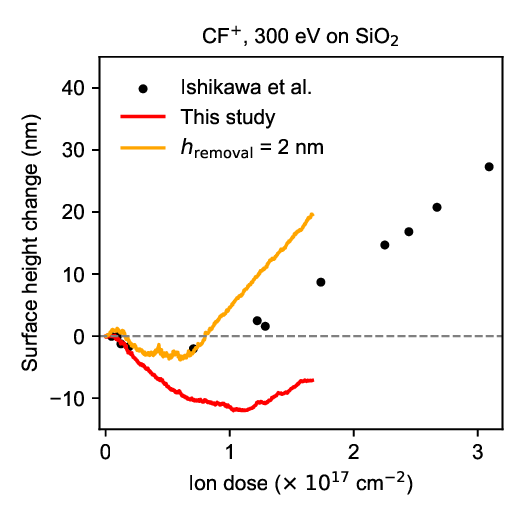}
\caption{Surface height change profile for CF$^+$ ions at 300 eV on a SiO$_2$ substrate, as a function of ion dose up to 1.66 × 10$^{17}$ cm$^{-2}$. 
The red line represents simulation results from this study, while the black circles indicate data from an experimental ion beam study \cite{ishikawa2003transitional}. 
The orange line corresponds to an etching MD simulation in which only byproducts located within 2 nm of the surface are removed.
} \label{fig:3_1_3_valid_transient_SiO2}
\end{figure}

\subsubsection{Surface height change}
In previous ion beam experiments, it was observed that at certain intermediate energies—not high enough for continuous etching but not low enough for immediate deposition—the surface initially undergoes etching but gradually transitions to deposition as the ion dose increases, the surface initially undergoes etching but gradually transitions to deposition as the ion dose increases \cite{ishikawa2003transitional,yanai2005mass}. 
To examine whether our simulations can reproduce this dose-dependent transition behavior, we perform simulations with an ion dose of 1 × 10$^{17}$ cm$^{-2}$ or higher. 
We define the slab height as the 98th percentile of atomic heights, and the thickness change as the difference from the initial slab height.

For the SiO$_2$ system, we use CF$^+$ ions at 300 eV, following the same conditions as in the experiments\cite{ishikawa2003transitional}. \autoref{fig:3_1_3_valid_md_snapshot} illustrates the evolution of the substrate surface as a function of ion dose. 
In the early stages, etching dominates and leads to the formation of a mixed layer at the surface, where substrate atoms combine with carbon and fluorine atoms. 
As the ion dose increases, a carbon-rich layer gradually accumulates above the mixed layer, eventually forming a distinct fluorocarbon film. 
Once this carbon layer becomes sufficiently thick, further ion bombardment no longer removes substrate atoms, marking a transition from etching to deposition. 
In experiments, such structural evolution is measured as changes in surface height: height profile thus reflects the combined effects of substrate etching, chemical mixing, and subsequent film deposition.

\autoref{fig:3_1_3_valid_transient_SiO2} shows that the thickness of the etched film before entering the deposition regime exceeds 10 nm, which is significantly larger than the experimental value of 2 nm. 
One possible reason for this overetching is the byproduct removal criterion: in our MD simulations, we assume that sufficient time elapses between ion impacts, allowing all stable byproducts to diffuse out of the slab regardless of their formation depth. 
However, in reality, when byproducts are generated deep within the slab, geometrical constraints, such as long diffusion paths, should be considered to construct a more realistic model. In our simulations, the distribution of byproduct removal heights peaks approximately 3 nm below the surface (Figure~S7), which can be too large distance to diffuse out from the surface.

To examine this effect, we repeat the MD simulation while restricting byproduct removal to within 2 nm (an arbitrary value) from the surface. 
Under this condition, the results more closely match experimental values, as shown in \autoref{fig:3_1_3_valid_transient_SiO2}. 
Although fixing the removal region can improve agreement, we opt to remove byproducts at all heights, as fitting for each ion species and energy is infeasible due to limited experimental data.
In Figure~S8, we also compare the evolution of surface composition (discussed in detail in the next section) when the removal height is applied.
For the SiO$_2$ system with CF$^+$ ions at 300 eV, the formation pattern of the carbon film shows slight differences, leading to minor variations in the early-stage evolution of Si and C contents.
Nevertheless, the overall trend remains consistent.
Once the system enters the carbon deposition stage, the fluorine content converges to a ratio of approximately 0.2 in both cases.

For the Si$_3$N$_4$ system, Figure~S9 compares the thickness evolution during etching simulations using CF$^+$ and CH$_2$F$^+$ ions, both at 1000 eV. Although experimental data \cite{ito2011hydrogen} are included for reference, the ion energies used in those measurements are not explicitly reported for this particular measurement. 
Still, the simulations qualitatively agree with the experimental trends, and we include the comparison to highlight this consistency.
Experimental observations highlight two key behaviors: both CF$^+$ and CH$_2$F$^+$ ions initially induce etching before transitioning to deposition as the ion dose increases, and CH$_2$F$^+$ maintains the etching phase for a longer duration than CF$^+$ \cite{ito2011hydrogen}. 
Our simulations reproduce these trends, showing a delayed transition to deposition for CH$_2$F$^+$. 
Additionally, although the two ions exhibit different deposition onset points, the subsequent growth rates are similar—a finding also consistent with experimental reports.

\subsubsection{Surface composition}
\begin{figure}[htbp]
\centering
\includegraphics[width=1.0\linewidth]{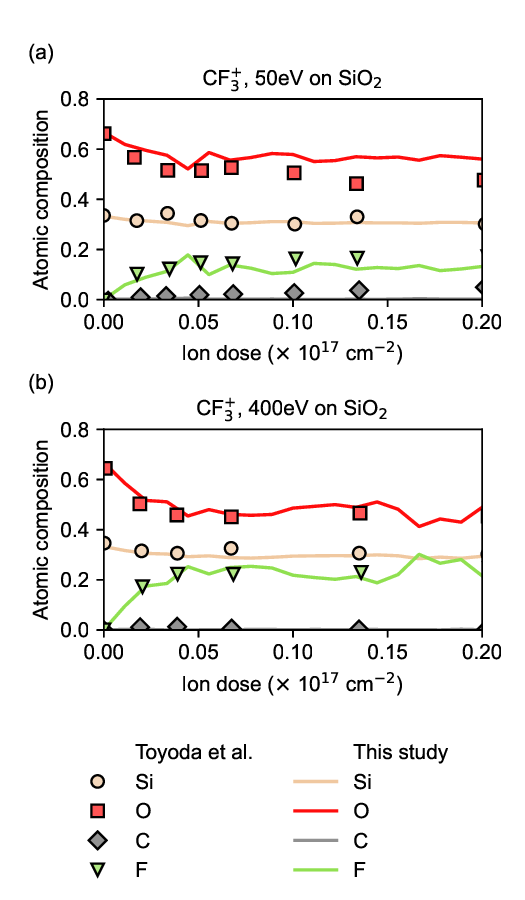}
\caption{Surface composition change in SiO$_2$ during etching with CF$_3^+$ ions \cite{toyoda2004beam}. Results are shown for (a) 50 eV and (b) 400 eV, demonstrating the energy dependence of fluorine incorporation and silicon depletion.} \label{fig:3_1_4_valid_surf_comp_SiO2}
\end{figure}

We also compare the surface composition changes during the etching process with experiments \cite{toyoda2004beam,ito2011hydrogen}.
The surface composition is evaluated within a 5 nm region from the topmost atom of the slab, which lies within the typical detection depth of XPS around 5--10 nm\cite{stevie2020introduction}.
For the SiO$_2$ system, we limit our analysis to the initial ion dose range up to 2 × 10$^{16}$ cm$^{-2}$, which corresponds to the point where the SiO$_2$ layer is fully etched in the experiment \cite{toyoda2004beam}. \autoref{fig:3_1_4_valid_surf_comp_SiO2} shows that the atomic composition changes and their convergence trends at the early dose stage for the CF$_3^+$ ion at 50 and 400 eV generally agree well with the experimental results.

\begin{figure}[htbp]
\centering
\includegraphics[width=1.0\linewidth]{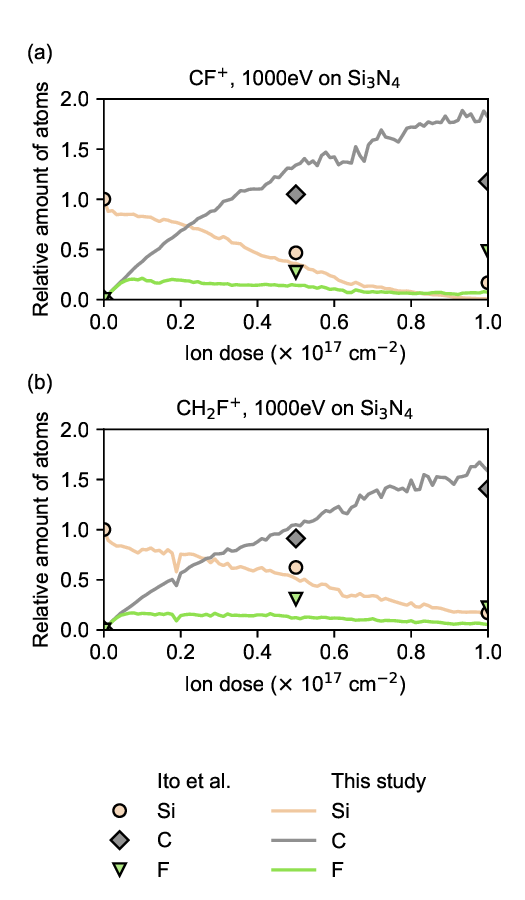}
\caption{Surface composition change in Si$_3$N$_4$ during etching with (a) CF$^+$ and (b) CH$_2$F$^+$ ions at 1000 eV \cite{ito2011hydrogen}.
Since the experimental data report elemental intensities in arbitrary units, we also present the relative amounts of Si, C, and F instead of absolute compositions. 
In our simulation results, the number of Si atoms before ion irradiation is set to 1, and all other values are normalized accordingly.
} \label{fig:3_1_4_valid_surf_comp_Si3N4}
\end{figure}

Similarly, \autoref{fig:3_1_4_valid_surf_comp_Si3N4} shows the surface composition changes for the Si$_3$N$_4$ system under CF$^+$ and CH$_2$F$^+$ ions at 1000 eV. Here, the ion dose is extended up to 1 $\times$ 10$^{17}$ cm$^{-2}$. Because a carbon film accumulates on Si$_3$N$_4$, the silicon fraction decreases while the carbon fraction increases, and both the trends and their relative magnitudes agree well with experimental observations.
These results further indicate that our NNPs accurately capture the overall surface dynamics during ion bombardment.

\subsection{Carbon film evolution across ion types and energies}

We now employ our NNPs to investigate the dynamic evolution of the substrate and fluorocarbon layer during plasma etching.
According to experimental ion beam studies, the dominant behavior—etching or deposition—varies depending on the ion type and incident energy \cite{ito2011hydrogen,yanai2005mass,karahashi2004etching}. 
Even under deposition conditions, the initial stage of ion exposure often results in surface etching. 
We aim to analyze these processes from an atomistic perspective and determine whether distinct behaviors emerge depending on ion type and energy. 
If such differences exist, we further examine their physical origins.

We define and quantify the following structural properties during the plasma etching process: thickness of the mixed layer and fluorocarbon layer.
To identify the mixed layer and fluorocarbon layers, we compute the atomic density profile along the z-axis and evaluate the relative concentration of C, H, and F atoms. 
Figure~S10 illustrates the method and parameters used to construct atomic density profiles, along with a representative example.
We define regions where the combined atomic ratio of C, H, and F exceeds 10\% as the mixed layer, and those exceeding 60\% as the film layer. 
These thresholds are selected to effectively separate substrate-modified regions from carbon-rich layers based on observed concentration profiles. 
Overlapping regions are considered part of the carbon-rich layer, and we assume that no mixed layer appear above the carbon layer.

\begin{figure}[htbp]
\centering
\includegraphics[width=1.0\linewidth]{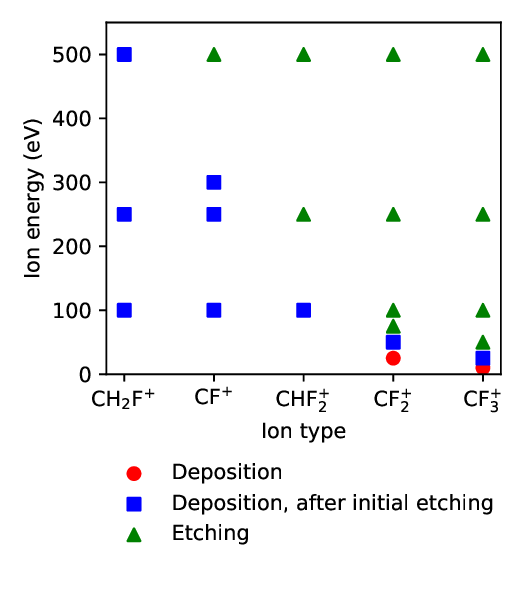}
\caption{Classification of etching behavior in the SiO$_2$ system based on ion type and energy.
The overall behavior is categorized into three regimes—deposition, deposition after initial etching, and etching—depending on the surface height profile and formation of carbon film layer.} \label{fig:3_2_1_regimeoverview}
\end{figure}

\begin{figure*}[htbp]
\centering
\includegraphics[width=1.0\linewidth]{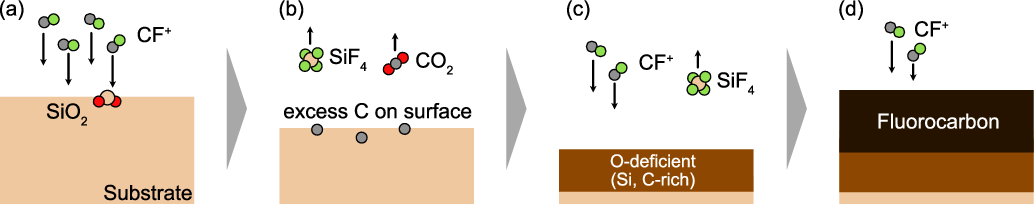}
\caption{Schematic illustration of fluorocarbon layer formation on the SiO$_2$ substrate under CF$^+$ ion irradiation.
(a) CF$^+$ ions impinge on the SiO$_2$ surface.
(b) Stable byproducts such as SiF$_4$ and CO$_2$ are formed, while excess carbon remains on the surface.
(c) As carbon accumulates, an oxygen-deficient mixed layer forms, primarily composed of Si and C. While Si can be removed as volatile SiF$_4$, carbon tends to remain on the surface.
(d) Once sufficient carbon has accumulated, a fluorocarbon layer begins to form. Subsequent CF$^+$ ions are incorporated into this layer, leading to gradual thickening of the fluorocarbon film.} \label{fig:3_2_1_mechanism}
\end{figure*}

\subsubsection{SiO$_2$}

Based on the evolution of surface height and the presence of fluorocarbon film formation, we classify the ion-induced surface evolution into three types: etching, deposition, and deposition after initial etching. \autoref{fig:3_2_1_regimeoverview} summarizes the classification of etching behavior by ion type and energy, while Figure~S11 presents the dose-dependent profiles of surface height and carbon film thickness for each ion species. 
When the ion energy is sufficiently high—above the threshold energy—etching occurs as expected in typical plasma etching processes. 
As shown in Figure~S11, the surface height decreases linearly with increasing ion dose, and only a minimal amount of fluorocarbon film remains on the surface. 
This regime typically reaches steady-state rapidly, within a dose of 0.2 $\times$ 10$^{17}$ cm$^{-2}$, which can be observed in the etch yield profiles in Figure~S5.

In contrast, when the ion energy is significantly lower than the threshold, etching does not occur and a fluorocarbon film is formed on the surface. 
This can be inferred from the parallel increase in both surface height and carbon film thickness as a function of ion dose. 
In this case, the fluorocarbon film grows via simple sticking of ions and radicals. The linear increase in surface height from the beginning of the dose suggests that the system quickly reaches a deposition regime.
Although we observed this behavior only for CF$_2^+$ and CF$_3^+$, it is likely a general phenomenon in the low-energy regime where the ion energy is insufficient to induce etching.

A more interesting regime appears when the ion energy is slightly below the threshold. 
In this case, etching initially occurs, reducing the surface height, but as the dose increases, a fluorocarbon film gradually forms on the surface, eventually leading to a transition into deposition. 
This behavior is clearly illustrated by the configuration changes with ion dose in \autoref{fig:3_1_3_valid_md_snapshot} and the corresponding surface height evolution in \autoref{fig:3_1_3_valid_transient_SiO2}.  
The energy range over which this etching-to-deposition transition occurs depends on the ion species. 
For CF$_2^+$ and CF$_3^+$, the transition window is relatively narrow, whereas for CF$^+$, CHF$_2^+$, and CH$_2$F$^+$, it is significantly broader.
Such differences in transition behavior across ion types have also been reported in previous ion beam experiments, particularly in a comparative study of CF$^+$ and CF$_2^+$ ions \cite{bello1994importance}.
In that study, CF$^+$ was found to form a Si-containing compound layer at 100 eV, indicating the onset of transition to deposition, whereas CF$_2^+$ exhibited clear etching behavior at the same energy. 
This contrast highlights that the transition window between etching and deposition is significantly broader for CF$^+$ than for CF$_2^+$.

This transition process is further illustrated in \autoref{fig:3_2_1_mechanism}.
Upon ion impact, a collision cascade breaks multiple bonds in the substrate, followed by dynamic recombination among fragments. 
In this process, carbon atoms preferentially react with oxygen, forming CO or CO$_2$, while fluorine atoms react with silicon to produce volatile SiF$_x$ species. 
Since oxygen can be removed by bonding with a single carbon atom, whereas removing silicon requires two to four fluorine atoms per atom, oxygen tends to be depleted more rapidly. 
This leads to the formation of an oxygen-deficient mixed layer, rich in Si and C, near the surface. 
When the F/C ratio of the incident ion is particularly low—as in the case of CF$^+$—excess carbon remains and accumulates in this region, contributing to surface thickening.
Once this mixed region is established, continued ion irradiation removes Si atoms near the surface, leading to the formation of a carbon film layer on top by remaining carbon atoms. 
As the film becomes sufficiently thick, it prevents incoming ions from reaching the underlying substrate, thereby marking the transition into a steady-state deposition regime.

\begin{figure}[htbp]
\centering
\includegraphics[width=1.0\linewidth]{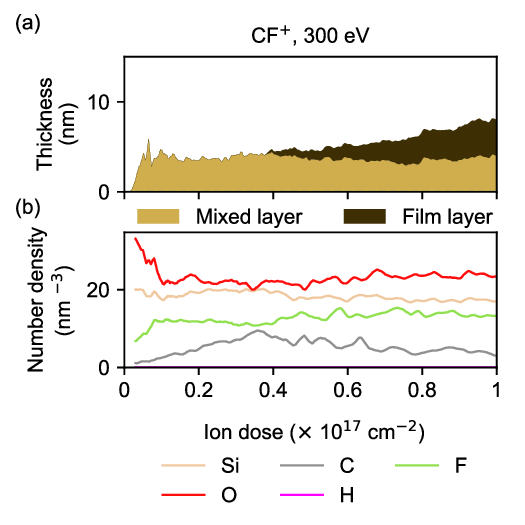}
\caption{Thickness and number density profile for CF$^+$ at 300 eV on the SiO$_2$ system.
(a) Thickness profiles of the mixed layer and carbon film layer as a function of ion dose.
(b) Number density profiles of atoms within the mixed layer region, providing insight into the structural composition of the interface.} \label{fig:3_2_1_CF_300}
\end{figure}

To quantitatively examine this process, we analyze the CF$^+$ case at 300 eV (\autoref{fig:3_1_3_valid_md_snapshot}) by plotting the thickness profiles of the mixed layer and carbon film, along with the number density of each atomic species within the mixed layer. \autoref{fig:3_2_1_CF_300} shows that after the formation of the mixed layer, its thickness remains nearly constant until the carbon film begins to grow. All values are smoothed by applying a moving average over an ion dose interval of 1 × 10$^{15}$ cm$^{-2}$.
During this stage, the oxygen content steadily decreases, while the silicon content remains relatively unchanged. This observation confirms the formation of an oxygen-deficient mixed layer near the surface.

We further investigated how the presence of hydrogen affects the formation of the mixed layer and carbon film. 
Experimentally, hydrogen is known to scavenge fluorine by forming HF, thereby promoting carbon accumulation on the surface \cite{ephrath1979selective,fukasawa1994high,horiike1995high,marra1997effect}.
Our MD simulations reveal consistent behavior, where hydrogen preferentially reacts with F and subsequently desorbs from the surface in the form of HF. 
This fluorine-scavenging behavior is reflected in the dose-dependent evolution of byproduct composition shown in Figure~S12.

This reduction in available fluorine limits etching reactions with silicon, making the etching process less effective. As a result, the surface height change is significantly suppressed, as seen in Figure~S11: for instance, CH$_2$F$^+$ at 250 eV shows almost no surface height reduction compared to CF$^+$ at the same energy.
To further understand this difference, Figure~S13 compares the structural evolution for the two ions. 
Although the thicknesses of the mixed layer and carbon film are similar, the mixed layer in the CH$_2$F$^+$ case contains more hydrogen and less fluorine. 
This suggests that in the CF$^+$ case, fluorine remains available in the subsurface region before the carbon film forms, sustaining Si–F reactions and allowing continued etching.
In contrast, CH$_2$F$^+$ lacks sufficient subsurface fluorine, and etching quickly diminishes despite comparable structural buildup.

\subsubsection{Si$_3$N$_4$}

\begin{figure*}[htbp]
\centering
\includegraphics[width=1.0\linewidth]{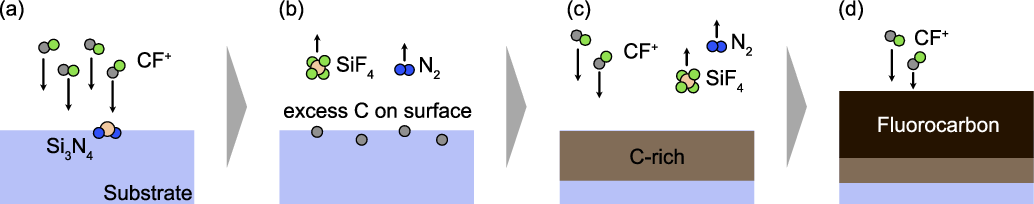}
\caption{Schematic illustration of fluorocarbon layer formation on the Si$_3$N$_4$ substrate under CF$^+$ ion irradiation.
(a) CF$^+$ ions impinge on the Si$_3$N$_4$ surface.  
(b) Stable byproducts are formed, while excess carbon remains on the surface. The main nitrogen-removal pathway is the formation of N$_2$, whereas no major byproducts are formed to remove carbon, leading to its accumulation on the surface.  
(c) Si is removed via reactions with fluorine, and N is eliminated independently via N$_2$ formation. As a result, both Si and N are depleted from the surface mixed layer, and a carbon-rich layer forms with carbon as the dominant component.  
(d) Once both Si and N are completely removed from the carbon-rich layer, it becomes a fluorocarbon layer.
} \label{fig:3_2_2_mechanism}
\end{figure*}

In SiO$_2$ system, we identified three distinct ion-induced surface evolution regimes: etching at high ion energies, deposition at low energies, and a transition from initial etching to subsequent deposition at intermediate energies. 
During the transition regime from etching to deposition, the preferential reaction of carbon with oxygen results in the formation of an oxygen-deficient mixed layer that eventually leads to the growth of fluorocarbon film. 
Next, we examine the behavior in Si$_3$N$_4$ system, which is known to exhibit reduced etching rate and and enhanced fluorocarbon film deposition compared to SiO$_2$ in experiments  \cite{yanai2005mass,schaepkens1999study,matsui2001relationship,matsui2001observation}. 
Unlike the SiO$_2$ case, our MD simulations show that fluorocarbon films in Si$_3$N$_4$ form at high ion doses irrespective of ion type and energy, with the exception of CF$_3^+$ at 1000 eV (Figure~S15). 
In this section, we discuss the mechanisms underlying this enhanced fluorocarbon deposition.


Based on the behavior of Si$_3$N$_4$ etching, we propose the carbon film formation mechanism shown in  \autoref{fig:3_2_2_mechanism}. 
Upon ion impact, a collision cascade breaks chemical bonds in the substrate, generating reactive fragments. 
Si atoms primarily react with fluorine to form volatile SiF$_x$ species, while N atoms are removed mainly as N$_2$ gas, whereas other nitrogen-containing gases such as HCN, FCN, and NH$_3$ appear only in trace amounts, as shown in Figure~S14. We note that N$_2$ has been reported as the major nitrogen-containing byproduct in experimental studies of Si$_3$N$_4$ plasma etching
\cite{clarke1985mass,field1988spectroscopic,winters1983etch}.
This contrasts with the SiO$_2$ system, where oxygen depletion is a prerequisite for carbon retention (\autoref{fig:3_2_1_mechanism}); in Si$_3$N$_4$, the reduced formation of volatile carbon–nitrogen compounds favors carbon accumulation regardless of the F/C ratio of the ions.
At higher ion doses, the mixed layer undergoes continued loss of Si and N atoms with concurrent C atom accumulation, eventually converting to dense fluorocarbon film.
Once formed, this film layer hinders the etching of Si and N atoms by blocking ion penetration, marking the onset of continuous fluorocarbon deposition.

\begin{figure}[htbp]
\centering
\includegraphics[width=1.0\linewidth]{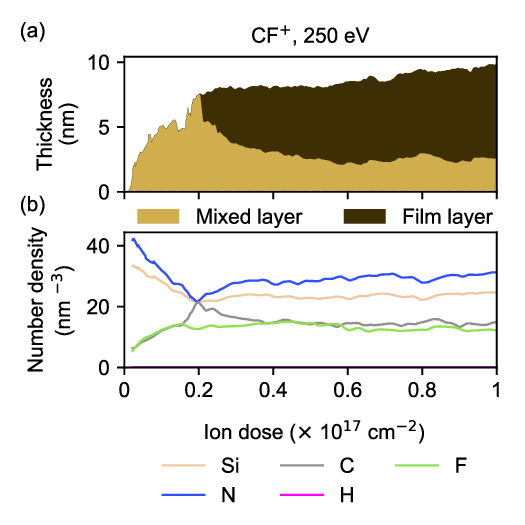}
\caption{Thickness and number density profiles under CF$^+$ ion bombardment at 250 eV on the Si$_3$N$_4$ system.
(a) Thickness profiles of the mixed layer and carbon film layer as a function of ion dose.  
(b) Number density profiles of atoms within the mixed layer region, revealing nitrogen depletion and early carbon incorporation near the interface.  
} \label{fig:3_2_2_CF_250_Si3N4}
\end{figure}

To further analyze the carbon film formation in Si$_3$N$_4$ system, we compare the thickness profiles of the mixed and fluorocarbon layers for CF$^+$ at 250 eV as shown in \autoref{fig:3_2_2_CF_250_Si3N4}. 
As the dose increases, the mixed layer is formed and thickened until a dose of 2 × 10$^{16}$ cm$^{-2}$. At higher dose, the carbon film layer forms with a decrease in the thickness of mixed layer, which then remains constant after 4 × 10$^{16}$ cm$^{-2}$ dose. This behavior differs from the SiO$_2$ system, where the thickness of mixed layer remains constant after fluorocarbon film formation at 4 × 10$^{16}$ cm$^{-2}$ dose.
In SiO$_2$ system, the mixed layer reaches a steady state where Si, O and F maintain constant number densities and the thickness remains constant, while the carbon number density increases. This implies that the carbon atoms in the mixed layer are easily removed from volatile byproducts such as CO, CO$_2$, the remaining carbon atoms are deposited on the surface of the mixed layer, as seen in \autoref{fig:3_1_3_valid_md_snapshot}. 
In contrast, the mixed layer in Si$_3$N$_4$ system reaches a different steady state from that of SiO$_2$, where Si and N number densities decrease while C, F number densities and thickness increase. From this profile, we expect that the carbon and fluorine atoms replace Si and N sites in the mixed layer due to the absence of facile process to remove carbon atoms as in the SiO$_2$ system, which results in a carbon-rich mixed layer at high doses and conversion to the mixed layer to fluorocarbon film. 
Furthermore, the atom number density profile reveals that nitrogen is removed more rapidly than silicon until the carbon film forms, as also observed in other cases, resulting in Si/N ratio in the mixed layer remaining lower than that of the bulk material.
This nitrogen depletion at the surface, attributed to preferential sputtering, has been reported in experimental studies of silicon nitride etching \cite{yanai2005mass,lee2000low,renaud2019two}.

To compare the influence of the number of fluorine atoms in the incident ions on fluorocarbon film formation, we analyze the atom number densities for CF$^+$, CF$_2^+$, and CF$_3^+$ ions at 250 eV, as shown in Figure~S16. 
In CF$^+$ case, N is reduced more rapidly than Si, whereas Si decreases abruptly in the CF$_2^+$ and CF$_3^+$ cases. Consequently, the Si/N ratio in the mixed layer is remains lower after carbon film formation compared to the CF$^+$ case. This demonstrates that more fluorine atoms can etch Si more efficiently, which is harder to remove by physical sputtering than N.
In the CF$_3^+$ case, while the Si and N number densities and F/C ratio in the mixed layer are similar to those of CF$_2^+$, the fluorocarbon film layers are less formed, indicating that the major variation between these ions is in the amount of film deposited. The formation of byproduct in Figure~S14 shows that the SiF$_4$ production is higher in CF$_3^+$ than in CF$_2^+$, while the N$_2$ production remains similar between them. These results confirm that the composition of incoming ions is a key determinant of etching behavior. Si-containing byproducts are more effectively removed with a higher F/C ratio in the ions, whereas nitrogen readily desorbs as N$_2$ molecules, making it less sensitive to variations in ion species.

For a given ion species, the surface height exhibits a greater reduction with increasing energy, as shown in Figure~S15. To assess the impact of ion energy on fluorocarbon film formation, we compare the results of etching MD simulations using CF$^+$ ions at 500, 750, and 1000 eV. 
Figure~S18 presents the atom number density profiles across energies and reveals that the Si/N ratio increases more dramatically at higher energies during the early stages of etching due to enhanced nitrogen removal. 
In addition, the production of N$_2$ byproducts is observed to increase with ion energy, as shown in Figure~S14. The amount of SiF$_4$ byproduct also increases with energy, but not as much as N$_2$; instead, the amount of SiF$_2$ byproduct increases with energy. 

Notably, even at the high energy of 1000 eV, the fluorocarbon film thickness continues to increase within the ion dose range initially considered. 
To determine whether this behavior reflects continuous carbon growth or merely a transient phase prior to reaching a steady state, we extend the MD simulations to a higher dose of 1.4 × 10$^{17}$ cm$^{-2}$ using CF$_3^+$ ions at 1000 eV.
Figure~S19 shows the surface height and carbon film thickness profiles: 
after a dose of 1.0 × 10$^{17}$ cm$^{-2}$, no further deposition is observed, and the film thickness plateaus, indicating the onset of a steady state.
Figure~S20 presents the number of atoms removed from the system either by sputtering (i.e., atoms ejected from the surface during MD) or by forming volatile byproducts that desorb after the ion incidence. 
The sputtering of Si and N atoms gradually decreases and eventually ceases, suggesting that the fluorocarbon film has grown thick enough to prevent further sputtering of the underlying substrate.
At the same time, the removal rates of C and F increase initially and then stabilize, indicating that carbon-containing species are ejected from the surface via chemical sputtering and subsequently removed as byproducts.
Therefore, these results demonstrate that, with a sufficiently high ion dose, continuous etching can proceed in the presence of a carbon film on the Si$_3$N$_4$ surface.

To investigate the role of hydrogen in the etching process, we compare the atom number densities between the CF$^+$ and CH$_2$F$^+$ cases in Figure~S17.
Similar to the behavior observed in the SiO$_2$ system, hydrogen scavenges fluorine by forming HF (Figure~S14), thereby reducing the number of available F atoms. This reduction in available fluorine suppresses etching reactions, leading to a higher number density of Si atoms within the mixed layer. Figure~S14 also indicates that nitrogen removal via the formation of NH$_3$, HCN, and related species is negligible compared to the dominant pathway of N$_2$ generation. Thus, the addition of hydrogen has little impact on the etching of nitrogen atoms.


Beyond the detailed comparison between SiO$_2$ and Si$_3$N$_4$, our simulations suggest a fundamental difference in how the fluorocarbon film evolves at the interface: 
while a persistent mixed layer remains beneath the carbon film in SiO$_2$, the corresponding layer in Si$_3$N$_4$ gradually transforms into a fluorocarbon layer as Si and N atoms are sputtered away and the layer becomes thinner. 
Although speculative, our findings point to the possibility that such interfacial dynamics—particularly the presence or absence of a stable mixed layer—can influence the controllability of etch depth in high-precision processes. 
These insights provide a potential direction for interpreting experimental results and refining process design strategies, such as precursor selection, film thickness control as ALE of SiO$_2$ and Si$_3$N$_4$ \cite{li2016fluorocarbon,metzler2016fluorocarbon,ishii2017atomic}.
\section{Conclusion}
In this study, we developed neural network potentials (NNPs) capable of modeling hydrofluorocarbon plasma etching processes in SiO$_2$ and Si$_3$N$_4$ systems. The training dataset encompassed a broad range of local environments—from substrates to fluorocarbon films—sampled using high-temperature MD trajectories at various densities, including vapor-to-surface sampling. Additional reference data included molecules, bulk, and slab structures. The NNPs were further refined through iterative training on trajectories from MD simulations of etching. The final models achieved good agreement with DFT results and were validated by comparing etch yields, surface height changes, and surface compositions with ion beam experiments.

Using these refined models, we analyzed system-dependent behaviors in reactive ion etching. SiO$_2$ exhibits a transition from etching to deposition within specific ion energy windows, driven by oxygen deficiency in the evolving mixed layer. In contrast, Si$_3$N$_4$ consistently forms fluorocarbon films under all tested conditions, attributed to the preferential sputtering of nitrogen as N$_2$ and the suppressed formation of carbon-nitrogen gas products. The framework enables quantitative predictions of atomistic surface modifications under plasma exposure, providing a foundation for integration with higher-level process modeling approaches such as kinetic Monte Carlo. Although the present study focuses on ion-induced effects and does not explicitly include radicals or neutrals, the established methodology can be extended to incorporate these species, enabling more comprehensive and predictive plasma process simulations.

\section{Acknowledgements}
This paper is the result of the research project supported by SK hynix Inc. The computations were carried out at Korea Institute of Science and Technology Information (KISTI) National Supercomputing Center (KSC-2023-CRE-0458) and at the Center for Advanced Computations (CAC) at Korea Institute for Advanced Study (KIAS).

\section{Data availability statement}
The codes used for simulation and data analysis are available at the following GitHub repository: 

https://github.com/HyungminAn/etchZBL

\printcredits

\bibliographystyle{ieeetr}

\bibliography{cas-refs.bib}

\end{document}